\documentclass{amsart}
\usepackage{graphicx}
\usepackage{caption}
\usepackage{subcaption}
\newtheorem{thm}{Theorem}[section]

\theoremstyle{definition}

\theoremstyle{remark}
\newtheorem{rem}[thm]{Remark}
\numberwithin{equation}{section}

\begin{document}

\author{V.M.Adamyan$^1$, I.Y.Popov$^2$, I.V.Blinova$^2$, V.V. Zavalniuk$^1$\\
	$^1$ Odessa I.I.Mechnikov National University, \\
	Dvoryanskaya str., 2, Odessa, 65082, Ukraine\\
	$^2$ ITMO University, Kronverkskiy, 49, \\
	Saint Petersburg, 197101, Russia \\
	popov1955@gmail.com }
\title{Simulation of detection and scattering of sound waves by the lateral line of a fish}
\date{ }

\begin{abstract}
An explicitly solvable model of sound sources detection and sound wave scattering by the lateral line of a fish or amphibian based on the theory of self-adjoint perturbations of the Laplace operator by arrays of additional boundary conditions at isolated points is proposed and discussed.
\end{abstract}

\maketitle
\section{Introduction}

Fishes and amphibians have special sensitive organs known as lateral line. It consists of a set of protuberances (neuromasts) located along a line on the head and body of the fish. These knobs can sit directly at the fish surface or can be inside canals connected with the environment (this structure is shown in Fig.
1). The lateral line of the fish consists of a fluid-filled tube lined with hair cells running along each side, just under the skin. This tube connects to the external environment via secondary fluid-filled tubules that branch off from the main tube and penetrate the skin at regular intervals. Water vibrations are transmitted from the secondary tubules to the main tube by sequential compression and expansion of fluid . These vibrations then jiggle the gelatinous domes of hair cells lining the main tube.

Sound is a multi-stage event that requires four components to occur: a source of vibration, a transmitting medium, a receiving detector, and an interpreting nervous system.Sound energy is carried by the oscillation of particles composing a transmitting medium. In the case of fishes, the transmitting medium is the water through which they swim. As for detection and interpreting, it is made by the lateral line and inner ears (many fish sensory biologists refer to the combination of inner ears and lateral lines as the acoustico-lateralis system). It is experimentally established, that fishes can hear sounds with frequencies ranging from about 10 Hertz to about 800 Hertz  \cite{CBH,M,Mar}. The upper range boundary differs for different fish species and in some cases extends to 4 kHz \cite{RDKP}. Various aspects of fish hearing (acoustical parameters of lateral line and swim bladder, description of sound receiver, etc.) are described in \cite{DGB,FH,Kal,MLP,LR,SB}.

In \cite{PP95,PP97} a physical explanation (based on resonance of incoming sound wave) is given for a mechanism of detection of the direction to sound source by means of the lateral line organ. In a few words, it is as follows. A system of
neuromasts can be approximately considered as a periodic system of open acoustical resonators. This system (more precisely, the wave equation
\begin{equation}\label{wave1}
\frac{\partial^{2}\psi}{\partial t^{2}}=A\psi
\end{equation}
for the velocity potential $\psi(\mathbf{r},t)$, where $A$ is the Laplace differential operator $-c^{2}\Delta$, $c$ is the speed of sound,  with the corresponding boundary conditions on the resonators) has characteristic features in the angular distribution of the intensity of sound waves scattered by the lateral line. If one considers a problem of plane wave scattering, then the result depends on the direction of the incoming and outgoing waves. Particularly with changes in the wave vector of the incident wave, sharp increases and decreases in pressure on the lateral line can be observed. The fish feels (through the neuromasts due to pressure surges) these changes and can determine the direction of the incoming wave by variation of the direction of it's own lateral line, i.e. of the fish body.

In this work, using these considerations as a starting point, we develop a solvable model that makes it possible to determine the excess pressure on the lateral line and describe the angular distribution of sound waves scattered from it in the field of an external sound source. The model is based on the theory of self-adjoint extensions of symmetric operators within which small biological resonators are replaced by point-like potentials. The obtained explicit expression of the Green function for this model allows us to compare the pressures on different sides of the fish and to find the differential cross section for sound scattering directly on the lateral line.

\begin{figure}
\includegraphics[scale=0.7]{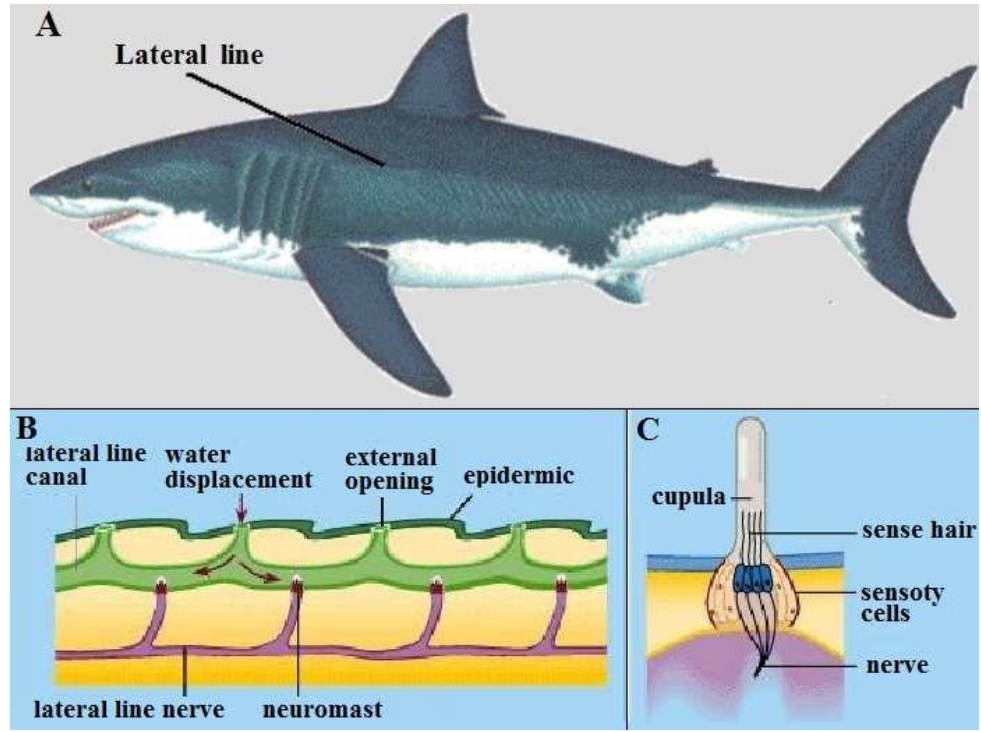}
\caption{The structure of the lateral line: A) Lateral line
position; B) Lateral line structure; C) Neuromast structure.}
\end{figure}

\section{Model description and Green function}

To construct a solvable model we replace small resonators (neuromasts) by point-like potentials and consider two parallel periodic chains of such potentials. The mathematical background is given by the theory of self-adjoint extensions of symmetric operators (see, e.g., \cite{AGHH,BEKS,P90,P92,PP93,PP93a}). To explain the procedure of extension, it is convenient to consider the simplest case of single point-like potential (at ${\bf r} =0$). One starts from the self-adjoint operator -- the Laplacian $-c^{2}\Delta$ in $L_2({\mathbb R}^3)$ with the domain $H^2({\mathbb R}^3)$, where $H^2$ is the Sobolev space $W^2_2$. Let us restrict the operator on the set of smooth functions vanishing at point ${\bf r} = 0$.
The closure of this restricted operator is a symmetric operator $A_0$ with deficiency indices (1,1). To construct a self-adjoint extension it is more convenient to deal with the corresponding restriction of the adjoint operator instead of the extension of the symmetric operator. There are several ways of extensions descriptions, e.g., boundary triplets method (\cite{BMN}, von Neumann formulas (\cite{BS}), Krein resolvent formula (\cite{PABP,PAB}). We will use here a variant of the second approach which allows one to present an element from the domain of the adjoint operator ($A_0^*$) in the following form:
$$
\psi({\bf r}) = \psi_0(x) +  C_0^{\psi} \frac{e^{is\left|{\bf
r}\right|}}{4\pi \left|{\bf r}\right|}, \quad s=\sqrt{q^{2}+i\varepsilon}, \; q=\frac{\omega}{c}, \; \varepsilon \geq 0, \; \mathfrak{Im}s\geq 0.
$$
To construct a self-adjoint extension, one calculate the boundary form for elements $\psi, \phi$ from the domain of the adjoint operator:
$$
(A^*\psi,\phi) - (\psi,A^* \phi) = \psi_0(0)\overline{C_0^{\phi}}
- \overline{\phi_0(0)}C_0^{\psi}.
$$
A self-adjoint extension is given by a linear relation which lead to annihilation of the form. Evidently, in this simple case the relation has the form
\begin{equation}\label{s-a_ext}
\psi_0(0) = b\cdot C_0^{\psi}, \; \mathfrak{Im}b = 0.
\end{equation}
Formally, the linear relation (\ref{s-a_ext}) takes the form of a "boundary condition" at the origin:
\begin{equation}\label{1.1.0}
\lim _{\left|{\bf r}\right|\rightarrow 0}\left[ \frac \partial
{\partial \left|{\bf r}\right|}-\beta\right] \left|{\bf r}\right|\Psi
=0,
\end{equation}
From now on we will assume that $\beta\geq 0$, since the the selfadjoint extensions of $A_0$ related to the acoustic problems should be non-negative operators.

Lets consider the spectral problem for two parallel chains of zero-range potentials (boundary conditions of the form (\ref{1.1.0})) in $\mathbb{R}^3$. We will assume that the second chain is shifted relative to the first by vector ${\bf g}=g_1{\bf e_1}+g_2{\bf e_2}$, where ${\bf	e_1},{\bf e_2}$ are unit vectors along the axis $X$ and  $Y$ respectively, and that the indicated zero-range potentials are located along the first and second chains at points  $n{\bf a} =na{\bf e_1}$ and $na {\bf e_1}+{\bf g}$, $n=0,\pm 1,\pm 2,\ldots $ respectively.

As in the previous case, starting from the self-adjoint operator -- the Laplacian in $L_2({\mathbb R}^3)$ with the domain $H^2({\mathbb R}^3)$ -- one can restrict this operator on the set of smooth functions vanishing at the indicated points of two chains. The closure of this restricted operator is a symmetric operator with infinite
deficiency indices. It has rich family of self-adjoint extensions. To construct a self-adjoint extension, one can take the corresponding adjoint operator. An element from its domain has the following form:
\begin{equation}\label{adjoint}
\psi({\bf r}) = \psi_0(x) + \sum_{n=-\infty }^{+\infty }\left\{
C_n^{\left(
1\right) }\frac{\exp \left[ is\left| {\bf r}-n{\bf a}%
\right| \right] }{\left| {\bf r}-n{\bf a}\right| }%
+C_n^{\left( 2\right) }\frac{\exp \left[ is\left| {\bf r}-n%
{\bf a}-{\bf g}\right| \right] }{\left| {\bf r}-n{\bf a}-{\bf
g}\right| }\right\},
\end{equation}
where $\psi_0$ belongs to the domain of the Friedrichs extension of the initial symmetric operator, $C_n^{\left( 1\right) }$ and $C_n^{\left( 2\right) }$ are some constants, such that the series (\ref{adjoint}) converges to a $L_{2}(\mathbb{R}^{3}$ -- function and $\mathfrak{Im}s>0$.
Choosing the extension  corresponding to the sum of the same as above zero-range potentials  at points $n\mathbf{a}, \,  n\mathbf{a}+\mathbf{g}, \, n=0,\pm 1,\pm 2,...$ that is, the extension determined by the boundary conditions
\begin{equation}\label{3a}
\underset {\rho^{\nu}_{n}\rightarrow 0}{\lim}\left[\frac{\partial}{\partial \rho^{\nu}_{n}}-\beta\right]\Psi =0, \quad    \nu=1,2 , \quad \rho^{1}_{n}=|\mathbf{r}-n\mathbf{a}|, \quad \quad \rho^{2}_{n}=|\mathbf{r}-n\mathbf{a}-\mathbf{g}|,
\end{equation}
yields a self-adjoint operator $A$ as required for the model in question.

The Green function for $A$ (the kernel of the integral operator $\left(A-s^{2}\cdot I\right)^{-1},\; \mathfrak{Im}s> 0, \; I\text{-unity operator}$) in general has
the form
\begin{equation}\label{3.1}\begin{array}{c}

G\left( {\bf r},{\bf r^{\prime }},s\right) =\frac 1{4\pi
}\frac{\exp \left[ is\left| {\bf r}-{\bf r^{\prime }}\right|
\right] }{\left| {\bf r}- {\bf r^{\prime }}\right| } \\

+\sum\limits_{n=-\infty }^{+\infty }\left\{ C_n^{\left(
1\right) }\frac{\exp \left[ is\left| {\bf r}-n{\bf a}%
\right| \right] }{\left| {\bf r}-n{\bf a}\right| }%
+C_n^{\left( 2\right) }\frac{\exp \left[ is\left| {\bf r}-n%
{\bf a}-{\bf g}\right| \right] }{\left| {\bf r}-n{\bf a}-{\bf
g}\right| }\right\}. \end{array}
\end{equation}
\begin{rem}\label{ax}
Note that the system of boundary conditions gives us a sufficient number of equations to uniquely determine the coefficients  $C_n^{\left( 1\right) },$ $C_n^{\left( 2\right) }$ using numerical methods. In general, the number of boundary conditions $N$ (equal to the total number of neuromasts) is finite, although it can be quite large. On the other hand, the length of the lateral line $\approx 2Na$, along which the neuromasts are located, is in reality much smaller than the characteristic distance $R$ between the fish and a sound source (or a sonar). Therefore, further, where this can not significantly affect the accuracy of calculations, we will use the limiting relations for $N\rightarrow\infty$. At the same time, at large distances $R$ from the lateral line,we shall use asymptotic expressions inherent in scattering problems.
\end{rem}

Setting
\begin{equation}\label{notat1}\begin{array}{c}

\xi _k^{\left( j\right) }=\sum_nC_n^{\left( j\right) }e^{ikna};
\quad j=1,2, \quad -\frac{\pi}{a}\leq k < \frac{\pi}{a},
 \\
\varphi _k^{(1)}\left( {\bf r}, s\right) =-\frac 1{4\pi }\sum_n\frac{\exp
	\left[ is\left| {\bf r}-n{\bf a}\right| +ikna\right] }{%
	\left| {\bf r}-n{\bf a}\right| }, \quad \varphi _k^{(2)}\left( {\bf r,s}\right) =-\frac 1{4\pi }\sum_n\frac{\exp
	\left[ is\left| {\bf r}-\mathbf{g}-n{\bf a}\right| +ikna\right] }{%
	\left| {\bf r}-\mathbf{g}-n{\bf a}\right| },\\
D\left( k,s\right) =\sum\limits_{n\neq 0}\frac{\exp \left[ is\left| n\right|
	a+ikna\right] }{\left| n\right| a}+is-\beta ,\quad
Q\left( k,s\right) =\sum\limits_n\frac{\exp \left[ is\left| n {\bf
		a}- {\bf g}\right| +ikna\right] }{\left| n{\bf a}- {\bf g}\right|},
\end{array}
\end{equation}
and using the boundary conditions (\ref{3a}) one can easily check that the functions $\xi _k^{\left( 1\right) },\xi _k^{\left( 2\right) }$ satisfy the following system of algebraic equations

\begin{equation}\label{3.2}
\begin{array}{c}
D\left( k,s\right) \xi _k^{\left( 1\right) }+Q\left( k,s\right)
\xi _k^{\left( 2\right) }=\varphi _k^{(1)}\left( {\bf r^{\prime
}},s\right),
\\
Q\left( -k,s\right)\xi _k^{\left( 1\right) }+D\left( k,s\right)
\xi _k^{\left( 2\right) }=\varphi _k^{(2)}\left( {\bf r^{\prime }},%
s\right).
\end{array}
\end{equation}

Taking into account the boundary conditions (\ref{3a}), expressions (\ref{3.5}), equalities (\ref{3.6}) and the inversion formula
\begin{equation}\label{inver}
C_{n}^{(j)}=\frac{a}{2\pi}\int\limits_{-\frac{\pi}{a}}^{\frac{\pi}{a}}e^{-ika}\xi_{k}^{(j)}dk
\end{equation}
we conclude that the following version of Krein's formula for the Green's function for $A$  is valid:
\begin{equation}\label{krein1}\begin{array}{c}

G\left( {\bf r},{\bf r^{\prime }},s^{2}\right) =\frac 1{4\pi
}\frac{\exp \left[ is\left| {\bf r}-{\bf r^{\prime }}\right|
	\right] }{\left| {\bf r}- {\bf r^{\prime }}\right| } -\frac{a}{2\pi}\int\limits_{-\frac{\pi}{a}}^{\frac{\pi}{a}}\sum\limits_{j,j^{\prime }=1}^{2} \gamma_{jj^{\prime }}(k)\varphi _{k}^{(j)}\left( {\bf r}, s\right)\varphi _{-k}^{(j^{\prime})}\left( {\bf r}^{\prime},s\right)dk,\\

\gamma_{11}(k,s)=\gamma_{22}(k,s)=\frac{D(k,s)}{W(k,s)}, \gamma_{12}(k,s)=\gamma_{21}(-k,s)=-\frac{Q(k,s)}{W(k,s)}, \\ W(k,s)=D^{2}(k,s)-Q(-k,s)Q(k,s) . \end{array}
\end{equation}

\section{Remote source detection and scattering }
Complex amplitude of steady-state oscillations of excess pressure $p(\mathbf{r},t)$ at an arbitrary point $\mathbf{r}$, induced by a point source $$A_{+}\delta(\mathbf{r}-\mathbf{R})e^{-i\omega t}, $$ which creates an excess variable pressure $p_{0}e^{-i\omega t}$  at a short distance $r_{0}$ from the source, coincides up to a constant factor with the Green function $G\left( {\bf r},{\bf R },q^{2}+i0\right)$,
\begin{equation}\label{pr1}
p(\mathbf{r},t)=4\pi r_{0}p_{0}G\left( {\bf r},{\bf R },q^{2}+i0\right).
\end{equation}
Within the framework of the model under consideration, expression (\ref{pr1}) is singular at the points mimicking neuromasts' positions on the lateral line. To get around this difficulty, assuming that $r_{0}$ in (\ref{pr1}) is significantly smaller than $a$, we will define the pressure at points $\mathbf{r}_{n}^{(1)}=n\mathbf{a},\mathbf{r}_{n}^{(2)}=n\mathbf{a}+\mathbf{g}, \; n=0,\pm 1, \pm 2,...$ as the average
\begin{equation}\label{average1}
p\left(\mathbf{r}_{n}^{(j)},t \right) =\frac{3}{4\pi r_{0}^{3}}\int_{\left|\mathbf{r}-\mathbf{r}_{n}^{(j)}\right|<r_{0}}p(\mathbf{r},t) d\mathbf{r}, \quad j=1,2.
\end{equation}
The Krein formula (\ref{krein1}) and explicit expressions for its entries (\ref{notat1},\ref{3.2}) enables us to make sure that
\begin{equation}\label{pr2}
p\left(\mathbf{r}_{n}^{(l)},t \right)\underset{r_{0}\rightarrow 0}{=}-\frac{3p_{0}a}{4\pi}\int\limits_{-\frac{\pi}{a}}^{\frac{\pi}{a}}\sum\limits_{j=1}^{2} \gamma_{lj}(k)\varphi _{-k}^{(j)}\left( {\bf r}^{\prime},s\right)dk,\cdot e^{-i\omega t}  \quad l=1,2.
\end{equation}
For simplicity, we have set above that $\mathbf{g}=g \mathbf{e}_{2}$. Taking into account that the sound absorption in water increases with frequency, we restrict ourselves also to cases when the  frequency of radiated sound satisfies the condition
\begin{equation}\label{freq1}
\omega<\frac{\pi\cdot c}{a}.
\end{equation}
Remember that \begin{equation}\label{asymp2}\begin{array}{c}
\underset{r\rightarrow\infty}{\lim}\underset{\varepsilon\downarrow 0}{\lim}\frac 1{4\pi
}\frac{\exp \left[ is\left| {\bf r}-{\bf R}\right|
	\right] }{\left| {\bf r}- {\bf R }\right| }=\frac{ 1}{4\pi
}\frac{e^{iqR}}{R}e^{-iq(\mathbf{b}\cdot\mathbf{r}}+O\left(\frac{1}{R^{2}} \right) , \\ s^{2}=\frac{\omega^{2}}{c^{2}}+i\varepsilon\equiv q^{2}+i\varepsilon, \quad \mathbf{b}=\frac{1}{R}\cdot\mathbf{R}=\left(\cos\theta_{X},\sin\theta_{X}\sin{\varphi},\cos\theta_{X}\cos{\varphi} \right),\end{array}
\end{equation}
and that
\begin{equation}\label{compl1}
\underset{N\rightarrow \infty}{\lim}\frac{a}{2\pi}\sum\limits_{n=-N}^{N}e^{i(ka-k^{\prime}a)n}=\sum\limits_{l=-\infty}^{\infty}\delta(ka-k^{\prime}a +2\pi l ).
\end{equation}

It follows from (\ref{pr2}) with account of  asymptotic expression (\ref{asymp2}), formula (\ref{compl1}) and the restriction (\ref{freq1}) that for a remote point source, the distance $R$ to which significantly exceeds the actual length of the lateral line, the excess pressures on two parallel parts of the lateral line are determined by the expressions
\begin{equation}\label{pr3}\begin{array}{c}
P^{(1)}(t)=-\frac{3p_{0}}{2R}e^{i(qR-\omega t)}\left[ \frac{1-e^{-iq(\mathbf{d}\cdot\mathbf{g})}}{D(q\cos{\theta_X},q^{2}+i0)-Q( q\cos{\theta_X},q^{2}+i0)}+\frac{1+e^{-iq(\mathbf{d}\cdot\mathbf{g})}}{D(q\cos{\theta_X},q^{2}+i0)+Q(q\cos{\theta_X},q^{2}+i0)}\right], \\ P^{(2)}(t)=-\frac{3p_{0}}{2R}e^{i(qR-\omega t)}\left[ \frac{1-e^{-iq(\mathbf{d}\cdot\mathbf{g})}}{D(q\cos{\theta_X},q^{2}+i0)-Q(q\cos{\theta_X},q^{2}+i0)}-\frac{1+e^{-iq(\mathbf{d}\cdot\mathbf{g})}}{D(q\cos{\theta_X},q^{2}+i0)+Q(q\cos{\theta_X},q^{2}+i0)}\right]
\end{array}
\end{equation}
The obtained expressions, in principle, make it possible to determine the direction of the radius vector drawn from a remote source to the object under consideration by the pressures $P^{(1)}(t), P^{(2)}(t)$.

In the case of scattering of a plane sound wave
\begin{equation}\label{free}
\psi_{0}(\mathbf{r},t)=\exp{i\left[(\mathbf{q}\cdot\mathbf{r})-\omega t\right]}, \quad \mathbf{q}=q\mathbf{b}\equiv\frac{\omega}{c}\mathbf{b}, \quad \mathbf{b}=\left(\cos{\theta_{X}},\cos{\theta_{Y}},\cos{\theta_{Z}} \right)
\end{equation}
by the considered system of point obstacles the solution of the wave equation (\ref{wave1}) satisfying the obvious condition
\begin{equation}\label{inf1}
\psi(\mathbf{r},t)=e^{-i\omega t}\left[e^{i\mathbf{q}\mathbf{r}}+u_{\omega}(\mathbf{r})\right], \quad   u_{\omega}(\mathbf{r})\underset{r\rightarrow \infty}{=} O\left( \frac{1}{r}\right)
\end{equation}
and the Sommerfeld radiation condition
\begin{equation}\label{inf2}
\underset{r\rightarrow \infty}{\lim}r\left[\frac{\partial u_{\omega}(\mathbf{r})}{\partial r}-iqu_{\omega}(\mathbf{r}) \right] =0
\end{equation}
are determined according to the Lippmann-Schwinger equation by the formula
\begin{equation}\label{lipsh}
\psi(\mathbf{r},t)=e^{-i\omega t}\underset{\varepsilon\downarrow 0}{\lim}-i\varepsilon\int_{\mathbb{R}^3}^{}G\left( {\bf r},{\bf r^{\prime }},q^{2}+i\varepsilon\right)e^{i\mathbf{q}\mathbf{r}^{\prime}}d\mathbf{r}^{\prime}.
\end{equation}
 (details can be found in \cite{MPS}). Note that solution (\ref{lipsh}) is nothing more than the result of the transformation of a plane monochromatic wave (\ref{free}) under the action of a wave operator $$ W(A,A_{0})=\underset{t\rightarrow -\infty}{s-\lim}\, e^{iAt}\cdot e^{-iA_{0}t}  $$ provided that the latter exists.

 Note that
 \begin{equation}\label{lipsh1}
\underset{\varepsilon\downarrow 0}{\lim}\frac{-i\varepsilon}{4\pi}\int_{\mathbb{R}^3}^{}\frac{\exp \left[ is\left| {\bf r}-{\bf r^{\prime }}\right|
	\right] }{\left| {\bf r}- {\bf r^{\prime }}\right| }e^{iq(\mathbf{b}\cdot\mathbf{r}^{\prime})}d\mathbf{r}^{\prime}=e^{iq(\mathbf{b}\cdot\mathbf{r})}, \quad s=\sqrt{q^{2}+i\varepsilon},  \quad |\mathbf{b}|=1.
 \end{equation}
Therefore, referring to the functions in (\ref{notat1}), in the limit $N\rightarrow \infty$ we have
\begin{equation}\label{ax2}\begin{array}{c}
\underset{\varepsilon\downarrow 0}{\lim}-i\varepsilon\int_{\mathbb{R}^3}^{}\varphi _{-k}^{(1)}\left( {\bf r}^{\prime},s\right)e^{iq(\mathbf{b}\cdot\mathbf{r}^{\prime})}d\mathbf{r}^{\prime}=\frac{2\pi}{a}\sum\limits_{l=-\infty}^{\infty}\delta\left( k-\frac{2\pi l}{a}-qb_{X}\right), \\ \underset{\varepsilon\downarrow 0}{\lim}-i\varepsilon\int_{\mathbb{R}^3}^{}\varphi _{-k}^{(2)}\left( {\bf r}^{\prime},s\right)e^{iq(\mathbf{b}\cdot\mathbf{r}^{\prime})}d\mathbf{r}^{\prime}=e^{iq\left( \mathbf{b}\mathbf{g}\right) }\frac{2\pi}{a}\sum\limits_{l=-\infty}^{\infty}\delta\left( k-\frac{2\pi l}{a}-qb_{X}\right).
\end{array}
\end{equation}
Hence, the part $u_{\omega}(\mathbf{r})$ of the velocity potential (\ref{inf1}) due to the scatterers by virtrue of (\ref{krein1}), (\ref{lipsh}) and (\ref{ax2}) in general is given by the formula
\begin{equation}\label{scatta1}\begin{array}{c}
u_{\omega}(\mathbf{r})=  -\sum\limits_{l=-\infty}^{\infty} \left\{ \left[ \gamma_{11}\left(qb_{X}+\frac{2\pi l}{a}\right)+\gamma_{12}\left(qb_{X}+\frac{2\pi l}{a}\right)e^{-i(\mathbf{q}\cdot\mathbf{g})}\right]
\varphi _{qb_{X}+\frac{2\pi l}{a}}^{(1)}\left( {\bf r}, s\right) \right. \\ \left. +\left[ \gamma_{21}\left(qb_{X}+\frac{2\pi l}{a}\right)+\gamma_{22}\left(qb_{X}+\frac{2\pi l}{a}\right)e^{-i(\mathbf{q}\cdot\mathbf{g})}\right]\varphi _{qb_{X}+\frac{2\pi l}{a}}^{(2)}\left( {\bf r}, s\right)\right\}  . \end{array}
\end{equation}
 Remembering again the asymptotic expression (\ref{asymp2}) we also see that
\begin{equation}\label{ax3}\begin{array}{c}
\underset{r\rightarrow\infty}{\lim}\underset{\varepsilon\downarrow 0}{\lim}\varphi _k^{(1)}\left( {\bf r}, s\right)=\frac{1}{2a}\frac{e^{iqr}}{r}\sum\limits_{l=-\infty}^{\infty}\delta\left(k-q\cos\vartheta_{X}- \frac{2\pi l}{a}\right),\\
\underset{r\rightarrow\infty}{\lim}\underset{\varepsilon\downarrow 0}{\lim}\varphi _k^{(2)}\left( {\bf r}, s\right)=\frac{1}{2a}\frac{e^{iqr}}{r}\cdot e^{-iq\left(\mathbf{d}\cdot\mathbf{g} \right) }\sum\limits_{l=-\infty}^{\infty}\delta\left(k-q\cos\vartheta_{X}- \frac{2\pi l}{a}\right).
\end{array}
\end{equation}
Using relations (\ref{scatta1},\ref{ax3}) we see that the scattering amplitude $f_{\omega}\left(\mathbf{d},\mathbf{b} \right)$, $$\underset{\varepsilon\downarrow 0}{\lim} \underset{r\rightarrow\infty}{=}u_{\omega}(\mathbf{r})=f_{\omega}\left(\mathbf{d},\mathbf{b} \right)\frac{e^{iqr}}{r}+O\left( \frac{1}{r^{2}}\right), $$ in the limit $N\rightarrow\infty$ has the form
\begin{equation}\label{scatta2}
\begin{array}{c}
f_{\omega}\left(\mathbf{d},\mathbf{b} \right)=  -\sum\limits_{l,l^{\prime}=-\infty}^{\infty} \frac{1}{2a} \left[ \gamma_{11}\left(qb_{X}+\frac{2\pi l}{a}\right)+\gamma_{12}\left(qb_{X}+\frac{2\pi l}{a}\right)e^{-i(\mathbf{q}\cdot\mathbf{g})}
  + \gamma_{21}\left(qb_{X}+\frac{2\pi l}{a}\right)e^{i(\mathbf{q}\cdot\mathbf{g})}\right. \\  \left. +\gamma_{22}\left(qb_{X}+\frac{2\pi l}{a}\right)e^{-i(\mathbf{q}\cdot\mathbf{g})}\right]\delta\left( q[b_{X}-d_{X}]-\frac{2\pi }{a}[l-l^{\prime}]\right)\chi_{\left[-\frac{\pi}{a},\frac{\pi}{a} \right]}\left(qb_{X}+\frac{2\pi l}{a}\right) , \\ \chi_{[\alpha_1,\alpha_2]}(x)=\left\lbrace \begin{array}{c}1, \quad x\in[\alpha_1,\alpha_2],\\ 0, \quad x\notin[\alpha_1,\alpha_2].\end{array} \right. \end{array}
\end{equation}

As follows from (\ref{scatta2}), scattering from the lateral line can be observed only when the projections of the wave vectors of the waves incident on this line and reflected from it satisfy the Bragg-type condition:
\begin{equation}\label{bragg}
q\left(\cos\theta_{X} -\cos\vartheta_{X}\right) =\frac{2\pi }{a}l, \quad l=0,\pm 1, \pm 2,... .
\end{equation}

If the sound wavelength is less than $2a$, then $b_{X}=d_{X}$ and
\begin{equation}\label{scatta3}\begin{array}{c}
f_{\omega}\left(\mathbf{d},\mathbf{b} \right)=  - \frac{1}{2} \left[ \gamma_{11}\left(q\cos\theta_{X},q\right)+\gamma_{12}\left(q\cos\theta_{X},q\right)e^{-iq(\mathbf{d}\cdot\mathbf{g})}
+\gamma_{21}\left(q\cos\theta_{X},q\right)e^{i(q\mathbf{d}\cdot\mathbf{g})}
\right. \\  \left. +\gamma_{22}\left(q\cos\theta_{X},q\right)\right]\delta\left( \cos\theta_{X}-\cos\vartheta_{X}\right). \end{array}
\end{equation}
Thus, for sound wavelengths greater than $2a$, that is if the condition (\ref{freq1}) holds, then sound scattering will occur while maintaining the projection of the wave vector on the direction parallel to the lateral line. Given the conservation of this component of the wave vector the scattering amplitude in the plane perpendicular to the lateral line as a function of $q=\frac{\omega}{c}$ and angles $\theta_{X}$, $$\varphi =\cos^{-1}\left(\frac{\left(\mathbf{d}\cdot\mathbf{b}\right)-\cos^{2}\theta_{X}}{\sin^{2}\theta_{X}} \right)    $$ is determined by the expression
\begin{equation}\label{scatta4}\begin{array}{c}
f_{\omega}\left(\vartheta_{X},\varphi \right)=- \frac{1}{2}\cdot\frac{2D\left(q\cos{\theta_{X}},q\right)-e^{-iq(\mathbf{d}\cdot\mathbf{g})}Q\left(q\cos{\theta_{X}},q\right)-e^{iq(\mathbf{d}\cdot\mathbf{g})}Q\left(-q\cos{\theta_{X}},q\right)}{D^{2}\left(q\cos{\theta_{X}},q-\right)-Q\left(q\cos{\theta_{X}},q\right)Q\left(-q\cos{\theta_{X}},q\right)}, \\D\left(q\cos{\theta_{X}},q\right)=\ln{\left( \frac{e^{-\beta a}}{2\left|\cos{qa }-\cos{\left(qa\cos{\theta_{X}} \right)}\right|}\right) } -i\pi, \\ Q\left( q\cos{\theta_{X}},q\right) =\sum\limits_n\frac{\exp \left[ iq\left| n {\bf
		a}- {\bf g}\right| +iq\cos{\theta_{X}}na\right] }{\left| n{\bf a}- {\bf g}\right|}.
\end{array}
\end{equation}

If it is permissible to neglect the displacement of the projections on the $X$-axis of the neuromasts of the left and right parts of the lateral line, for example, if $a\ll|\mathbf{g}|$ and $qa\ll 1$, then due to the equality $Q(k,s) = Q(-k,s)$, the last expression is simplified and takes the form
\begin{equation}\label{scatta5}\begin{array}{c}
f_{\omega}\left(\vartheta_{X},\varphi \right)=- \cdot\frac{D\left(q\cos{\theta_{X}},q\right)-\cos(\mathbf{q}\cdot\mathbf{g})Q\left(q\cos{\theta_{X}},q\right)}{D^{2}\left(q\cos{\theta_{X}},q\right)-Q^{2}\left(q\cos{\theta_{X}},q\right)}\\
=- \frac{1}{2}\left[\frac{1-\cos(\mathbf{q}\cdot\mathbf{g})}{D\left(q\cos{\theta_{X}},q\right)-Q\left(q\cos{\theta_{X}},q\right)}+ \frac{1+\cos(\mathbf{q}\cdot\mathbf{g})}{D\left(q\cos{\theta_{X}},q\right)+Q\left(q\cos{\theta_{X}},q\right)}\right].
\end{array}
\end{equation}

\section{Discussion}

The starting point for the development of the model proposed in this work was the mechanism of sound detection by the lateral line of fishes proposed in  \cite {PP95}, \cite{PP97}, where the corresponding analysis was generally qualitative. In the present paper we give a quantitative instrument for simulation, which makes it possible to quantitatively distinguish the pressures caused by an external source on two sides of the lateral line and to explicitly determine the differential cross section of sound scattering on the lateral line. They are the key issues in the describing of any acoustical system.

The expressions obtained for the excess pressures on the lateral line on different sides of the fish in the presence of a remote point source emitting sound vibrations with a certain frequency demonstrate a pronounced dependence of the amplitudes and phases of these pressures both on the frequency of emitted vibrations and on the angles $\theta_{X}$ and $\phi$, which determine (at constant distance from the source) orientation of  the unit vector $\mathbf{b}=\left(\cos{\theta_{X}},\sin{\theta_{X}}\cos{\phi},\sin{\theta_{X}}\sin{\phi}\right)$ of the straight line drawn through the distant source and the lateral line.

\begin{figure}
    \centering
    \begin{subfigure}[c]{0.4\textwidth}
        \includegraphics[width=\textwidth]{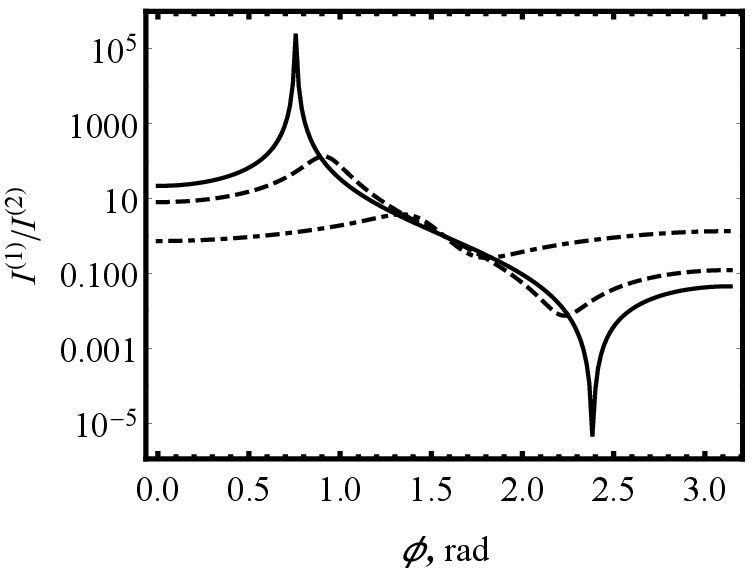}
        \caption{}
        \label{fig_qusiboundresX4}
    \end{subfigure}
    \hspace{1 cm}
    \begin{subfigure}[c]{0.4\textwidth}
        \includegraphics[width=\textwidth]{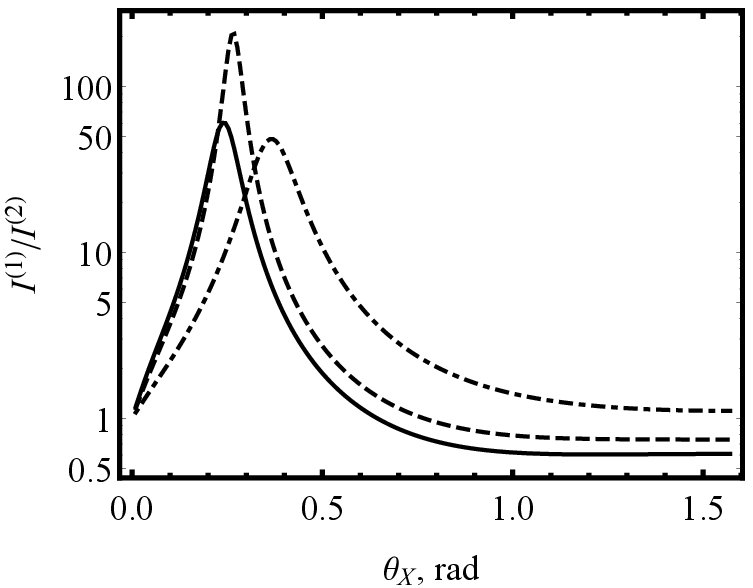}
        \caption{}
        \label{fig_qusiboundresX1}
    \end{subfigure}
    \\
    \begin{subfigure}[c]{0.4\textwidth}
        \includegraphics[width=\textwidth]{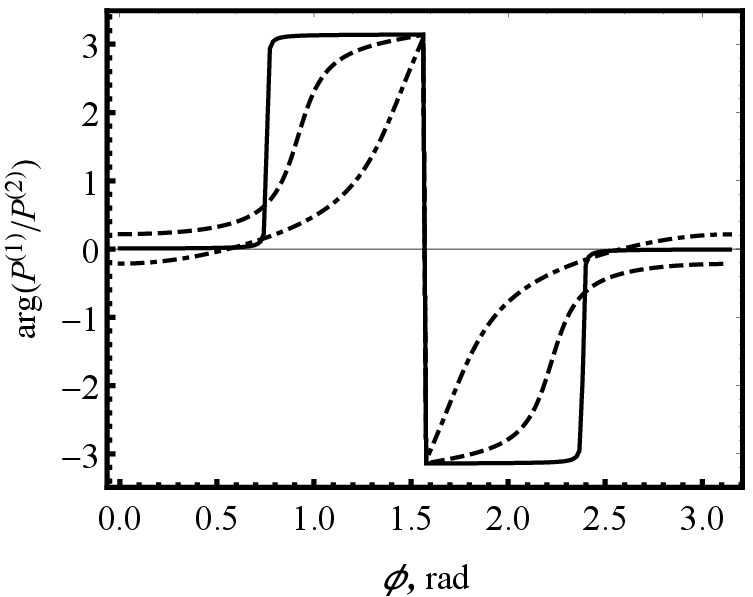}
        \caption{}
        \label{fig_qusiboundresX4}
    \end{subfigure}
    \hspace{1 cm}
    \begin{subfigure}[c]{0.4\textwidth}
        \includegraphics[width=\textwidth]{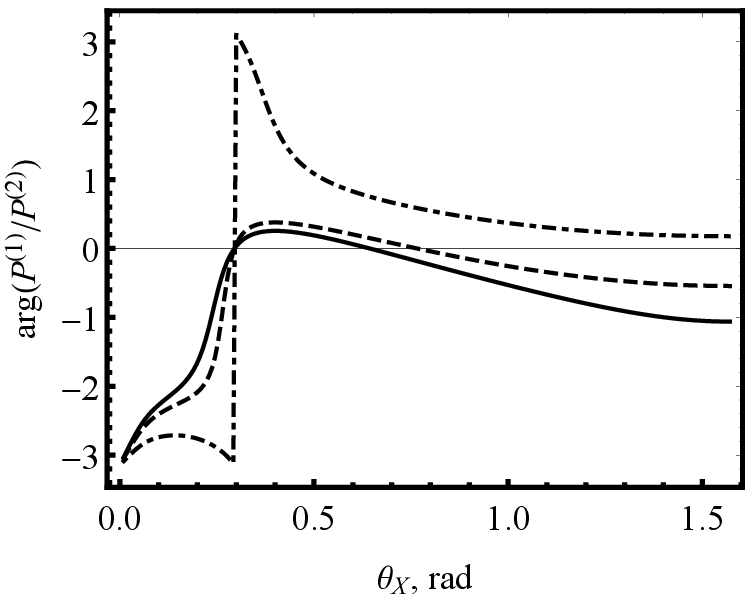}
        \caption{}
        \label{fig_qusiboundresX1}
    \end{subfigure}
    \caption{The ratio of sound intensities $\frac{I^{(2)}}{I^{(1)}}=\frac{\|P^{(2)}(t)\|^2}{\| P^{(1)}(t)\|^2}$ and the phase difference of pressures $P^{(1)}(t),P^{(2)}(t)$ on the parallel components of the lateral line, calculated by formula (\ref{pr3}) for the frequency corresponding to $qa = \frac{\pi}{6}$: A)  $\frac{I^{(2)}}{I^{(1)}}$ as function of $\theta_{X}$ for different values of $\phi$; B)  $\frac{I^{(2)}}{I^{(1)}}$  as function of $\phi$ for different values of $\theta_{X}$; C) $\arg\left(\frac{P^{(2)}(t)}{P^{(1)}(t)} \right)$ as function of $\theta_{X}$ for different values of $\phi$; D)$\arg\left(\frac{P^{(2)}(t)}{P^{(1)}(t)}\right)$ as function of $\phi$ for different values of $\theta_{X}$.}
\end{figure}

Dependence on the angles increases with increasing frequency and manifests itself strongly for the phase difference and amplitudes quotient for fluctuations of excess pressures on different halves of the lateral line along sides of the fish.

In the case of sound scattering on a lateral line within the framework of the considered model, the projections of  wave vectors of the incident and scattered sound waves onto the lateral line, as expected from symmetry considerations, are preserved. In this case, the dependence of  scattering amplitude on the frequency,  projection of the wave vector of the incident (scattered) wave to the lateral line, and the angle between  projections of wave vectors of the incident and scattered onto the plane perpendicular to the lateral line also becomes the more pronounced, the higher frequency.

\begin{figure}
    \centering
    \begin{subfigure}[b]{0.4\textwidth}
        \includegraphics[width=\textwidth]{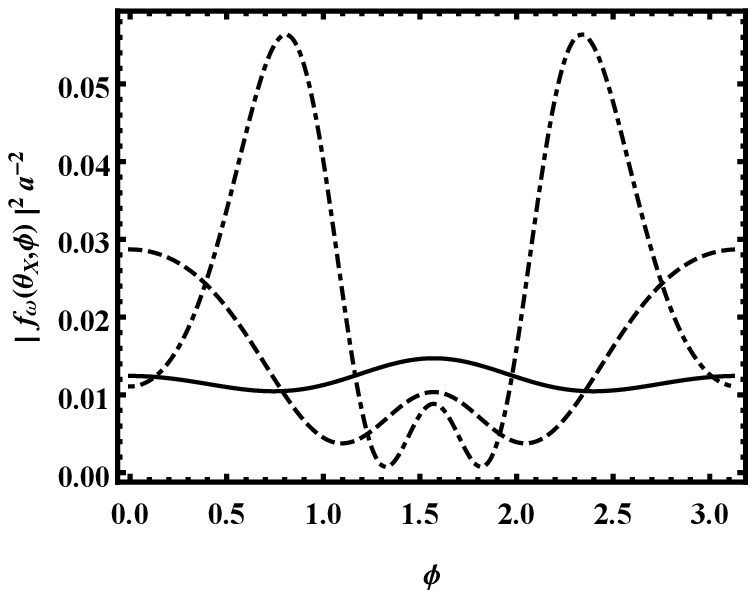}
        \caption{}
        \label{fig_qusiboundresX4}
    \end{subfigure}
    \hspace{1 cm}
    \begin{subfigure}[b]{0.4\textwidth}
        \includegraphics[width=\textwidth]{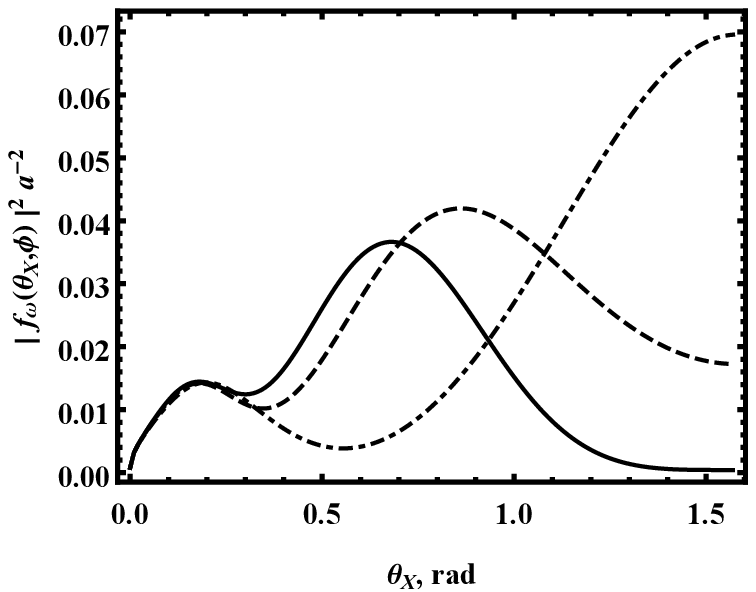}
        \caption{}
        \label{fig_qusiboundresX1}
    \end{subfigure}
    \caption{Dependence of the differential scattering cross section $\left| f_{\omega}\left(\theta_{X},\phi \right)\right|^{2}$ on the angle $\theta_{X}$ between the wave vector of the incident (scattered) sound wave and the direction vector of the lateral line and on the angle $\phi$ between projections of the wave vectors of the incident and scattered waves on the plane perpendicular to the lateral line, calculated by formula (\ref{scatta5}) for the frequency of the sound wave also corresponding to $qa = \frac{\pi}{6}$.}
\end{figure}

The described model can be simply improved. Instead of point potentials, one can use resonators with a point window approximation (see \cite {P90, P92}) or point potentials with an internal structure \cite {Pa}. For fishes with lateral lines of the "surface" type, the Green function for free space should be replaced by the Green function for two half-spaces separated by a slab with different speed of sound, and for fishes with lateral lines of the "channel type" - by the Green function for three-dimensional space with a potential with periodic internal structure, concentrated on two parallel lines. All these changes make it possible to significantly expand the scope of the model under consideration.

\section*{Acknowledgements}

This work was partially financially supported by the grant
0115U003208  of Ministry of Education and Science of Ukraine, by
the Government of the Russian Federation (grant 08-08), by grant
16-11-10330 of Russian Science Foundation. The authors thank Dr.
R.Fricke for useful advice.

\end{document}